\documentstyle[11pt,paspconf]{article}

\begin{document}

\title{Two Species of the Local Group Dwarf Spheroidals}
\author{
  Tsutomu T. Takeuchi $^{1}$ \&   
  Hiroyuki Hirashita$^{1}$}
\affil{
  Dept. of Astronomy, Faculty of Science, Kyoto University,
  Sakyo-ku, Kyoto 606-8502, JAPAN}

\altaffiltext{1}{
  Research Fellow of the Japan Society for the Promotion of
  Science}

\begin{abstract}
We analyzed 10 dwarf spheroidal galaxies (dSphs) in the Local Group, 
and found two distinct sequences on the $M_{\rm vir} 
/ L$ -- $M_{\rm vir}$ plane: $M_{\rm vir} / L \propto 
M_{\rm vir}^{1.6}$ for $M_{\rm vir} < 10^8\; M_\odot$ whereas 
$M_{\rm vir} / L \simeq {\rm const.}$ for $M_{\rm vir} > 
10^8\; M_\odot$ ($M_{\rm vir}$ and $L$ are the 
virial mass and the total luminosity of a dSph, respectively). 
We interpret the disconuity as the threshold for the gas in dSphs to be
blown away by successive supernovae. 
We succeeded to give a quantitative 
explanation of the discontinuity mass of $10^8M_\odot$ as blow-away 
condition. 
We further derived the above relation for the low-mass dSphs, 
assuming that the initial star formation 
rate of the dSphs is proportional to the inverse of the 
cooling time. 
The relation of high-mass dSphs is also explained along 
with the same consideration, with the condition that the gas cannot 
been blown away. 
\end{abstract}


\section{Discovery : $M_{\rm vir}/L$--$L$ Relation}

\setcounter{footnote}{1}
First of all, we present the relation between virial mass
$M_{\rm vir}$ and total $V$-luminosity $L$ of 10 Local
Group dSphs in Fig. 1. 
We divide these galaxies into
two groups; the lower-mass group (Group L; Draco, Carina,
Ursa Minor, Sextans, Sculptor, and Leo II) and the
higher-mass group (Group H; Fornax, NGC 147, NGC 185, and
NGC 205)\footnote{The figure is a revised one omitting the data point 
of Leo I because its value remains uncertain (Mateo et al. 1998).
Our conclusion is not seriously affected by 
this revision. The quantities of other dSphs are summarized in Hirashita 
et al. (1998b).}. 
The virial mass of the former group is less than $10^8M_\odot$, 
while that of the latter is more than $10^8M_\odot$.
We see that $M_{\rm vir}/L$ of Group H is almost
constant, though the number of samples is small. 
As for Group H, the constant $M_{\rm vir}/L$ may be due 
to the inefficiency of escape of SN-heated gas
(Dekel \& Silk 1986; Mac Low \& Ferrara 1998 (MF98); 
Ferrara \& Tolstoy 1998 (FT98)), since the
potential of dark matter (DM) is deep.
This is discussed in the next section. On the other hand, for
galaxies belonging to Group L, the gas 
easily escapes out of them once their gas is heated by SNe
(Saito 1979a) and OB-star radiation, because of
their shallow gravitational potentials. We discuss this point
later in \S 3. 

\begin{figure}
\vspace{8cm}
\caption{The relation between $M_{\rm vir}$ and $M_{\rm vir}/L$ 
  of the Local Group dwarf spheroidal galaxies.}
\end{figure}

\section{Blow-away Condition for Proto-dwarfs}

As discussed by Saito (1979a), SN-driven winds blow the
gas in low-mass proto-galaxies away because of their
shallow potential wells.
Recent numerical simulations of SN-driven wind at initial
starburst of low-mass galaxies (MF98; FT98) showed that if their
gas mass is more than $10^7M_\odot$, the gas ejection
efficiency is very low.
This mass corresponds to $M_{\rm vir}
=10^8M_\odot$ if the mass fraction of baryon is 0.1.
This value agrees with the separation line of the two groups, 
H and L.
This fact is physically interpreted as follows. Galaxies
with $M_{\rm vir}>10^8M_\odot$ form stars whose total mass is
proportional to the initial gas mass, which leads to the
constant mass-to-light ratio (Dekel \& Silk 1986). On the other
hand, galaxies with $M_{\rm vir}<10^8M_\odot$
blow the gas away soon after the formation of the
first-generation stars:
If a little fraction of gas becomes stars,
the SNe or UV-radiation heating resulting from these stars are enough to 
blow away
the rest of the gas (Saito 1979a; Nath \& Chiba 1995). Thus, the
members in the low-mass category tend to have little baryonic
matter (i.e., higher $M_{\rm vir}/L$).

\section{Low Mass Dwarf Spheroidals}

The virial mass and total luminosity
of the lower-mass group (Group L) in Fig. 1 satisfy the
relation, 
\begin{eqnarray}
M_{\rm vir}/{L}\propto M_{\rm vir}^{1.6}.
\label{obs}
\end{eqnarray}
In this section, we will derive this relation by considering
the physical processes in the formation epoch of dwarf galaxies.
For details, see Hirashita et al. (1998a).
The presence of DM in dSphs is indicated by observations of
velocity dispersions (e.g., Mateo et al. 1993; discussion based on
the structure formation is given in Flin 1998). 
Thus, it is reasonable to consider the gas
collapse in DM potential to form proto-dSphs. 
We assume that the distribution of DM is not affected by baryon, since
the mass fraction of baryon is much lower than DM.

We assume that the initial star formation rate (SFR) of a dSph is
determined by the cooling time $t_{\rm cool}$ 
($t_{\rm cool}\ga t_{\rm grav}$,
where $t_{\rm grav}$ is the free-fall time $\simeq 1/
\sqrt{G\rho_{\rm DM}}\sim 10^7$ yr).
We, here, note that the free-fall time is determined by the DM
mass density $\rho_{\rm DM}$, while the cooling time is determined
by the gas density.
The cooling time $t_{\rm cool}$ is
determined by the density and temperature of the gas as
\begin{eqnarray}
t_{\rm cool} \propto T/\rho_{\rm gas}\Lambda (T), \label{cool}
\end{eqnarray}
where $\Lambda (T)$ is the cooling function. 
Since the virial temperature of such a low mass galaxy is 
much lower than $10^6$ K, the cooling is dominated by H and He 
recombination (Rees \& Ostriker 1977): 
$\Lambda (T)\stackrel{\propto}{\sim}T^{-1/2}$
(Peacock \& Heavens 1990). 
From the above assumption that the SFR is
determined by the cooling
timescale, we obtain the following expression for the SFR
($\dot{M}$):
\begin{eqnarray}
\dot{M}\propto t_{\rm cool}^{-1}\propto
\rho_{\rm gas}{T^{-3/2}}. \label{sfr}
\end{eqnarray}
The temperature in the quasistatic collapse phase is
determined by virial temperature (Rees \& Ostriker 1977).
Thus, the following expression for the temperature is satisfied:
\begin{eqnarray}
T \propto M_{\rm vir}/R \; ,\label{virial}
\end{eqnarray}
where $R$ is the
typical size of the DM distribution.
Using the scaling relation of virial mass and size
 (Saito 1979b; Nath \& Chiba 1995);
$R\propto M_{\rm vir}^{0.55}$,
we obtain from relations (\ref{sfr}) and (\ref{virial})
\begin{eqnarray}
\dot{M}\propto M_{\rm vir}^{-1.33}, \label{scale}
\end{eqnarray}
where we assume that the initial mass ratio of gas to DM is
constant (i.e., $\rho_{\rm gas}\propto M_{\rm vir}/R^3\propto
M_{\rm vir}^{-0.65}$.)
In normal galaxies, the SF is stopped in the epoch of onset of
galactic wind, when the thermal energy produced by SNe
becomes equal to the binding energy of the galaxy (Arimoto
\& Yoshii 1987). 
However, in dwarf galaxies belonging to the
low-mass group, the heating by UV photons from first-generation OB
stars is large enough to supply the thermal energy equal to the
gravitational potential ($\sim 10$ OB stars $\sim 1$ OB
association). 
Thus the remaining gas in a low-mass dwarf
evaporates from the system soon after the formation of
first-generation stars.
We estimate the SF-terminating time by using
the crossing time of the wind generated
in an OB association. The velocity of the wind is estimated by the
sound speed $c_{\rm s}$ of $10^6$ K, which is a typical temperature
of the heated gas. The crossing time $t_{\rm cross}$ is estimated
as $t_{\rm cross}\simeq R/c_{\rm s}$. 
If we keep $c_{\rm s}$
constant, the following expression for the
$t_{\rm cross}$ is obtained by using the above scaling law
$R\propto M_{\rm vir}^{0.55}$:
\begin{eqnarray}
t_{\rm cross}\propto M_{\rm vir}^{0.55}.\label{cross}
\end{eqnarray}
We note that the typical value for $t_{\rm cross}$ becomes
$10^7$ yr (for $R=1$ kpc and $c_{\rm s}=100$ km s$^{-1}$), which is
shorter than the cooling time.
Thus, the formation of the second-generation stars is difficult.
The total mass of the stars, $M_*$, is estimated by
$\dot{M}t_{\rm cross}$. Thus, from relations (\ref{scale}) and
(\ref{cross}), we obtain $M_*\propto M_{\rm vir}^{-0.78}$.
This relation means that the mass-to-light ratio
is scaled as
\begin{eqnarray}
M_{\rm vir}/{L}\propto M_{\rm vir}^{1.78},
\label{theor}
\end{eqnarray}
where $L$ is considered to be proportional to $M_*$.

Comparing the observed relation (\ref{obs}) with the theoretical
prediction (\ref{theor}),
we see that the two agree well in the range of the observational
error. We note that it is difficult to obtain an absolute
value for the mass of the formed stars, since the star formation
efficiency is uncertain. Thus, we here only calculate the
scaling relation.

\acknowledgments

We are grateful to Drs.
A. Ferrara, M.-M. Mac Low and E. Tolstoy for their kindly sending us
their preprints and helpful comments on
the starburst-driven mass loss. 
We also thank Drs. K. Freeman, M. G. Lee, and S. van den Bergh for the 
useful comments on Fig. 1.
We acknowledge the Research Fellowship of the Japan Society for the
Promotion of Science for Young Scientists.

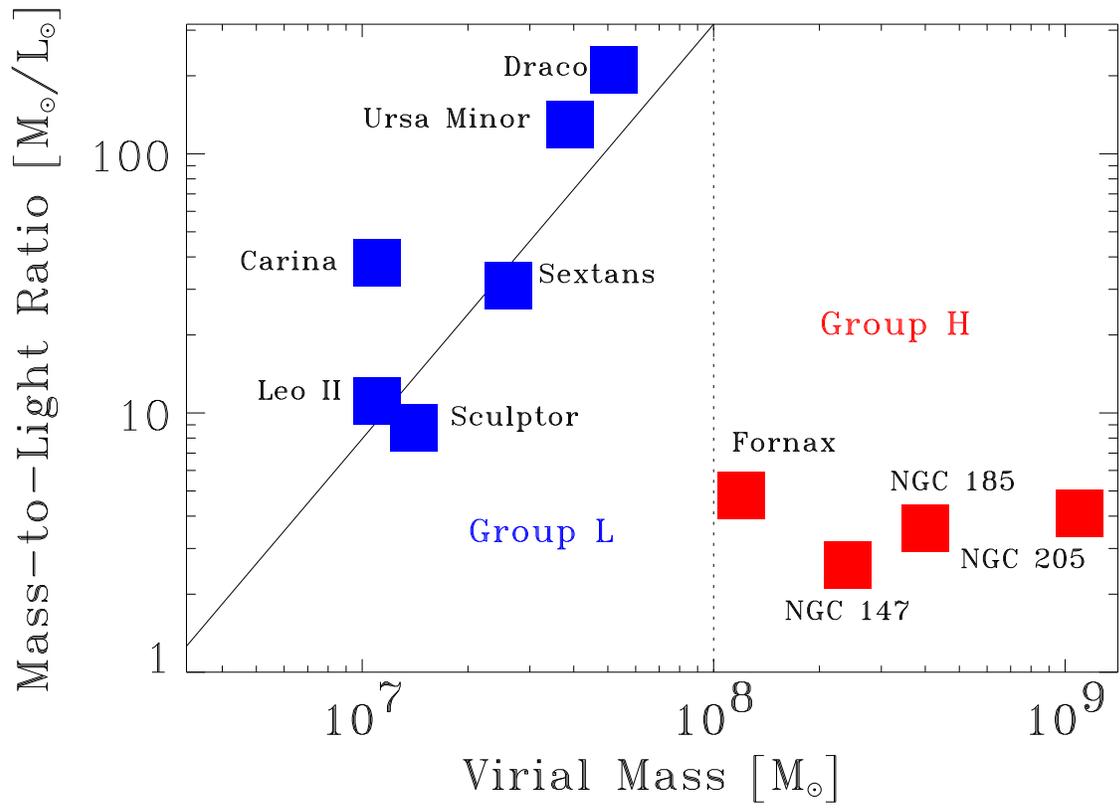
\end{document}